\documentclass[aps,prd,secnumarabic,superscriptaddress,amsmath,amssymb,nofootinbib,notitlepage,twocolumn]{revtex4-1}

\usepackage{graphics}      
\usepackage{graphicx}      
\usepackage{longtable}     
\usepackage{calrsfs}
\usepackage{float}
\usepackage{eucal}
\usepackage{bbold}
\usepackage{color, soul}
\usepackage{amsmath}
\usepackage{multirow}
\usepackage{verbatim}
\usepackage{url}           
\usepackage{bm}            
\usepackage{dsfont}
\usepackage{verbatim}
\usepackage{microtype}
\usepackage{hyperref}
\usepackage{array}
\usepackage{makecell}
\usepackage[dvipsnames]{xcolor}

\DeclareSymbolFont{matha}{OML}{txmi}{m}{it}
\DeclareMathSymbol{\varv}{\mathord}{matha}{118}

\renewcommand{\rm}{\mathrm}

\begin{document}
\title{Gravitational time dilation in extended quantum systems: the case of light clocks in Schwarzschild spacetime}
\author{Tupac Bravo}
\email{t.bravo21188@gmail.com}
\affiliation{University of Vienna, Faculty of Physics, Boltzmanngasse 5, 1090 Vienna, Austria}
\author{Dennis R\"atzel}
\email{dennis.raetzel@physik.hu-berlin.de}
\affiliation{Institut f\"ur Physik, Humboldt-Universit\"at zu Berlin, Newtonstraße 15, 12489 Berlin, Germany}
\affiliation{ZARM, Unversität Bremen, Am Fallturm 2, 28359 Bremen, Germany}
\author{Ivette Fuentes}
\email{I.Fuentes-Guridi@soton.ac.uk}
\thanks{Previously known as Fuentes-Guridi and Fuentes-Schuller.}
\email[Author to whom correspondence should be addressed: ]{I.Fuentes-Guridi@soton.ac.uk}
  \affiliation{School of Physics and Astronomy,
University of Southampton,
Southampton SO17 1BJ,
United Kingdom}
\date{\today}

\begin{abstract}

The precision of optical atomic clocks is approaching a regime where they resolve gravitational time dilation on smaller scales than their own extensions. Hence, an accurate description of quantum clocks has to take their spatial extension into account. In this article, as a first step towards a fully relativistic description of extended quantum clocks, we investigate a quantized version of Einstein's light clock fixed at a constant distance from a large massive object like the Earth. The model consists of a quantum light field in a one-dimensional cavity in Schwarzschild spacetime, where the distance between the mirrors is fixed by a rigid rod. By comparing a vertical and a horizontal clock, we propose an operational way to define the clock time when the clock resolves gravitational time dilation on scales smaller than its extension. In particular, we show that the time measured by the vertical light clock is equivalent to the proper time defined at its center. We also derive fundamental bounds on the precision of these clocks for measurements of proper time and the Schwarzschild radius. 

\end{abstract}

\maketitle

\section{Introduction}

 Atomic clocks are the most precise systems available to measure time. In recent years, they have become precise enough to resolve gravitational time dilation on the scale of millimeters \cite{bothwell2022resolving,zheng2022differential}. At these scales, it becomes important to take into account that atomic clocks are not point-like but spatially extended and that also their spatial degrees of freedom have quantum properties. For example, the atoms that are part of the clock experiment reported on in \cite{bothwell2022resolving}
 are extended in a region that is larger than the length scale on which time dilation is observed. In \cite{bothwell2022resolving}, this resolution of gravitational time dilation across the atomic sample was the effect that was to be measured. In other situations, it may be a systematic effect that has to be compensated for. In general, the concept of proper time is not anymore applicable as a property of the complete system and can only be used for distinct parts that are smaller than the scale of resolution.  The situation is even more complicated if the atoms within the sample at different heights are entangled. Entanglement can improve precision by reaching the Heisenberg limit.  However, if time runs at different rates for atoms entangled at different heights, the usual notion of clock time looses meaning and it becomes necessary to redefine it. Since gravitational time dilation is a relativistic effect, a consistent description of the clock at the interface of general relativity and quantum mechanics has to be applied. 
 
In this paper, we work towards a covariant description of quantum clocks by studying gravitational time dilation in a quantized light clock using quantum field theory in curved spacetime. The techniques developed here could be applied in future works to the atomic case which is of more practical interest.  The classical light clock model has been introduced in general relativity by Einstein; light pulses bounce back and forth between two mirrors. Equivalently, one can consider the time evolution of light modes in the optical resonator defined by the end mirrors. These modes can then also be quantized. In \cite{Lindkvist:2015,Lock:2019} such a model of a quantized version of a light clock was introduced to study the difference between quantum and classical light clocks moving in flat spacetime under non-uniform acceleration. The quantized clock model is underpinned by relativistic quantum field theory and therefore, the equations of motion are Lorentz-invariant and quantum probabilities, given by the Klein-Gordon inner product, are conserved. This model of a quantized light clock can be used to learn how quantum effects might affect time dilation in the presence of a black hole horizon, and how the curvature of Earth affects the precision of the clock. 

In this paper, we use the model of \cite{Lindkvist:2015,Lock_2019} to study the precision of a quantum light clock undergoing uniform acceleration in a Schwarzschild metric. This situation corresponds to a clock held at a fixed distance from the center of a spherically symmetric compact object, such as the Earth, or from the horizon of an eternal black hole. The time evolution of light clock modes in Schwarzschild metric was studied in \cite{Lock_2019}. For a thorough approach to the description of the fundamental frequency of a trapped field in a cavity in terms of the curvature of the underlying spacetime, see \cite{Raetzelfreq:2018}. Here we show how the acceleration affects the ticking of the clock and its precision. In particular, we deviate from the strategy used in previous analyses \cite{Lindkvist:2015,Lock_2019} by following a local operational approach to clock comparison: Instead of comparing a local clock with a clock at space-like infinity, we consider two local clocks that are differently oriented with respect to the direction of acceleration. Then, a comparison of time measurements by the two clocks reveals the combined effect of the finite extension of the clocks and gravitational time dilation. The imprint of the Schwarzschild radius can then, in principle, be used for its measurement. 

\begin{figure}
\includegraphics[width=\linewidth]{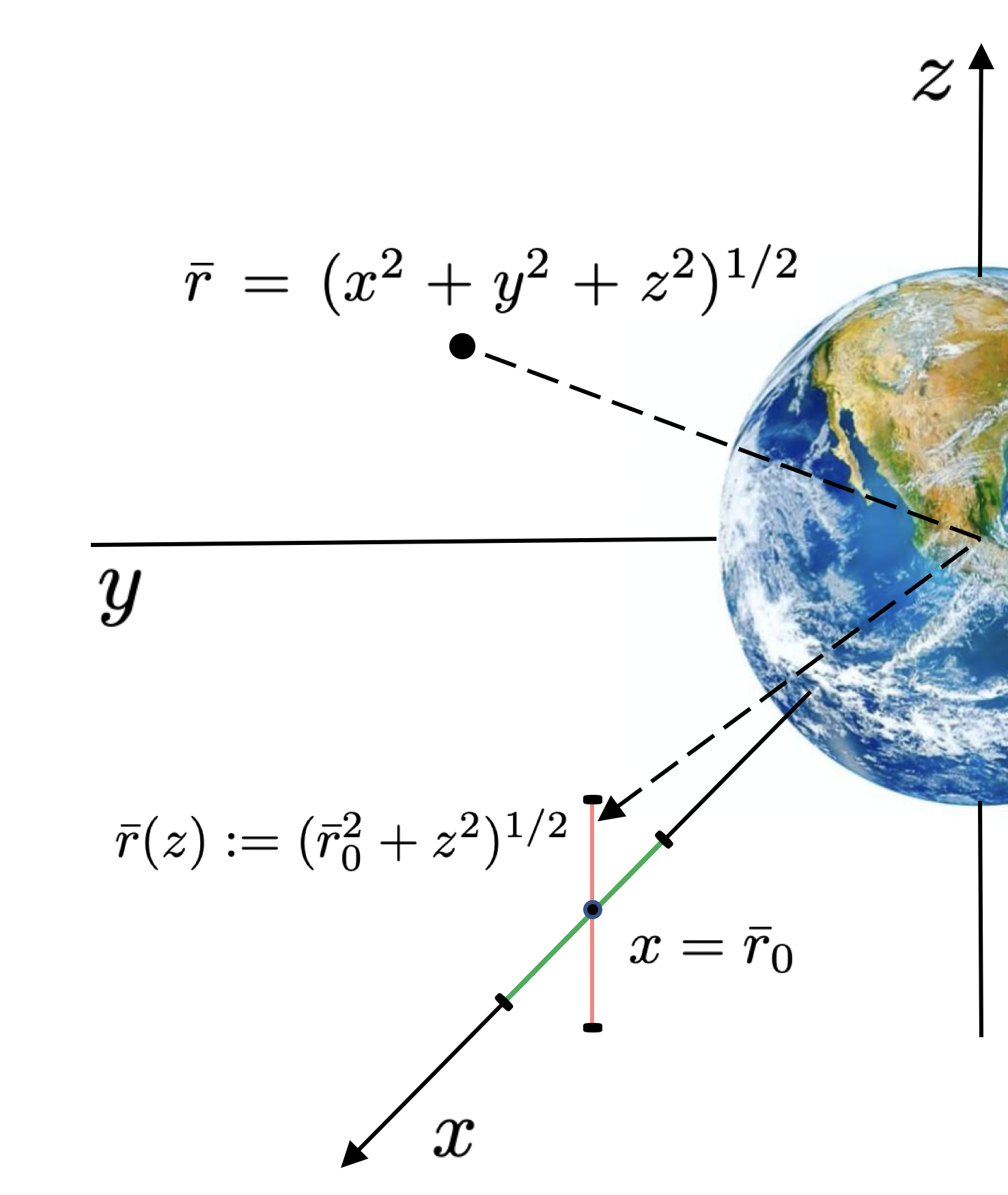}
\caption{Coordinate system used to quantize the field in horizontal (red) and vertical (green) cavities.}
\label{fig:coordinates}
\end{figure}

This paper is organised as follows: We use the Schwarzschild metric to model the spacetime of the Earth and of an eternal black hole. In Sec. \ref{sec:hovering_cavity}, we introduce an approximate version of the Schwarzschild metric appropriate for our purposes. In Sec. \ref{sec:quantization}, we introduce the clock model by describing the quantum field of a $1$-dimensional perfectly (Born) rigid cavity at a fixed distance from the central massive body, for example, the Earth. We consider two different cavity orientations, one vertical (aligned with the radial coordinate) and the other horizontal (perpendicular to the radial coordinate).  In section \ref{sec:clocks}, we show how the acceleration necessary to cancel the gravitational effect of the central massive object affects the frequency spectrum of the cavities. Applying quantum metrology methods, we provide bounds on the precision for the estimation of proper time measured by the cavity mode for coherent and squeezed clock states. Furthermore, we discuss how the time difference shown by the differently oriented clocks may be used for the estimation of the acceleration parameter, and thus, of the spacetime's Schwarzschild radius. We provide expressions for the fundamental precision limit of such a measurement. In Sec. \ref{sec:discussion}, we discuss the effect of finite rigidity of the cavity and use state-of-the-art parameters to give a realistic estimate of the precision at which the quantized light clocks can measure proper time, in principle. This is followed by our conclusions in Sec. \ref{sec:conclusions}.

We use the metric signature $(-,+,+,+)$ in $3+1$-dimensions and $(-,+)$ in 1+1 dimensions. The symbol Tp is used for transposition and we use bold font for matrices.

\section{Approximation of the Schwarzschild spacetime for small curvature}\label{sec:hovering_cavity} 
The clock model that will be analyzed consists of a quantized electromagnetic field confined in a rectangular cavity located at a fixed distance from a spherical body with mass $M$. In this section we will derive the effective spacetime metric assuming that the curvature is small  within the cavity. 
The spacetime surrounding a spherically symmetric compact object is well described by the Schwarzschild metric
\begin{equation} \label{eq:schwarzschild}
\boldsymbol{g} = \text{diag} \left(-f(r), \frac{1}{f(r)}, r^2, r^2 \sin^2\vartheta \right)\,,
\end{equation}
where we have used spherical Schwarzschild coordinates $x^\mu=(x^0,r,\vartheta,\phi)$, with $x^0=ct$, $r>0$ and $f(r) = 1-r_{\textrm{S}}/r$. The Schwarzschild radius is given by $r_{\textrm{S}}:= 2GM/c^2$, where $G$ is Newton's gravitational constant and $c$ the speed of light in the vacuum. The metric defines infinitesimal proper distances between points in spacetime through the line-element $ds^2$ given by $ds^2:=g_{\mu\nu}dx^\mu\,dx^\nu$.\footnote{We assume Einstein's summation convention on repeated indices.} 

Since the cavities are held at a fixed distance from the surface of the central object, it is more convenient to write the Schwarzschild metric in isotropic Cartesian coordinates $x^{\mu\prime}=(x^0,x,y,z)$, which are related to the spherical coordinates through $r = (1+r_{\textrm{S}}/4\bar{r})^2\bar{r}$, and $\bar{r}=(x^2+y^2+z^2)^{1/2}$. This transformation is time independent, therefore, it does not introduce any additional acceleration when fixing the coordinate position in the new coordinate system. The metric \eqref{eq:schwarzschild} in isotropic Cartesian coordinates reads
\begin{equation}\label{eq:SSiso}
	\boldsymbol{g}= \left(1 + \frac{r_{\textrm{S}}}{4\bar{r}}\right)^4\,\rm{diag}\left( - \frac{(1 - \frac{r_{\textrm{S}}}{4\bar{r}})^2}{(1 + \frac{r_{\textrm{S}}}{4\bar{r}})^6} ,1,1,1\right)\,.
\end{equation}
The spacetime curvature within the cavity is small when the distance from the cavity to the mass center is much larger than the Schwarzschild radius, i.e., $r_{\textrm{S}}/\bar{r}\ll1$. In this case we can expand the metric using a Taylor series around this parameter and neglect all contributions of higher than second order. The expansion of the metric is 
\begin{equation}\label{eq:schwarzschildapp}
\boldsymbol{g}=\boldsymbol{\eta} + \frac{r_{\textrm{S}}}{\bar{r}}\,\mathds{1} + \frac{r_{\textrm{S}}^2}{2\bar{r}^2}\rm{diag}\left(-1,\frac{3}{4},\frac{3}{4},\frac{3}{4}\right)\,,
\end{equation}
where $\boldsymbol{\eta}=\rm{diag}(-1,1,1,1)$ is flat Minkowski metric and $\mathds{1}=\rm{diag}(1,1,1,1)$ is the identity matrix. The relativistic corrections to Newtonian gravity are explicit in this expansion. The first two terms correspond to the Newtonian limit. They can be derived directly from the linearized Einstein equations and the energy momentum tensor for a non-relativistic point particle in the Lorenz gauge. However, the terms of order $(r_{\textrm{S}}/\bar{r})^2$ are purely relativistic. Considering a cavity held in a laboratory on the surface of the Earth, $r_{\textrm{S}}/\bar{r}\ll1$ is a good approximation since the Schwarzschild radius is $r_{\textrm{S}}\sim8.7\,\rm{mm}$ and the distance to the centre of the Earth is $6.37\times10^6\,\rm{m}$. The Earth's rotation and any deviations from spherical symmetry can be safely ignored for small cavities.

\section{\label{sec:quantization}Field Quantization of cavities in the Schwarzschild spacetime}

\begin{figure}
\includegraphics[width=\linewidth]{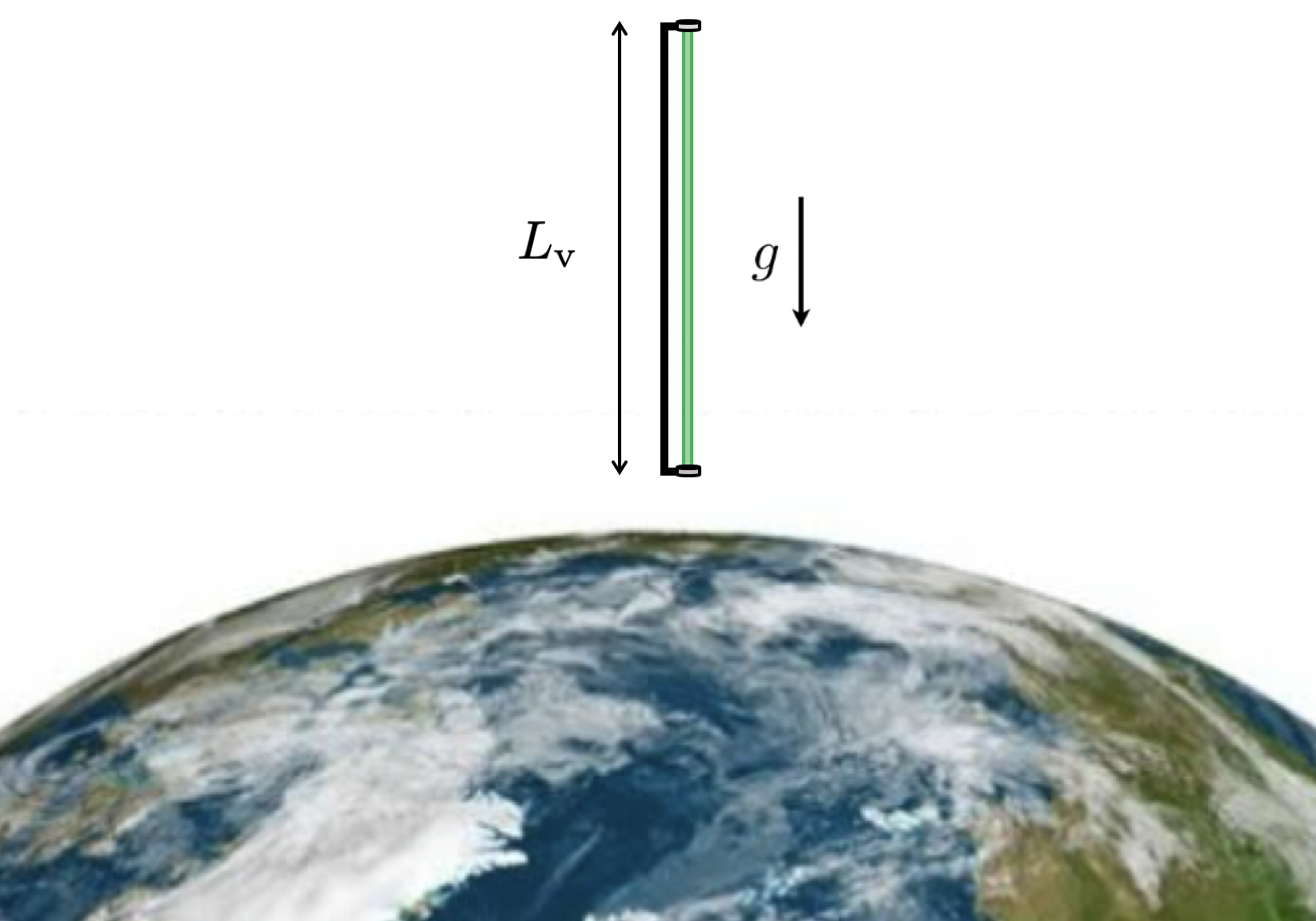}
\caption{Vertical light clock in the gravitational field of the Earth. The mirrors are held together by a $L_{\textrm{v}}=20$cm rigid rod.}
\label{fig: setup:second}
\end{figure}

Cavities that are held at a constant radius from the center of the spherically symmetric massive object do not follow geodesics. The cavities experience a proper acceleration pointing outwards in the radial direction. Two cavity dimensions are assumed to be much smaller than the third one so that the system can be effectively treated in one spatial dimension. The cavity mirrors are perfectly reflective and held together by a rigid rod. We assume the rigidity to be infinite. In section \ref{sec:discussion}, we will consider rods with a finite rigidity. We will show that the acceleration affects differently the time evolution of the light field in the case that the cavity is aligned (vertical) or perpendicular (horizontal) to the radial coordinate. In both cases there is a redshift that depends on the radial location of the cavity. For small horizontal cavities, the spacetime is flat along their length. Therefore, the horizontal cavity will be used as a reference clock in the laboratory frame. The vertical case (see figure \ref{fig: setup:second}) has an additional redshift term depending on its location with respect to the horizontal cavity. 

The proper length of the rod holding the mirrors placed at positions $x^\mu_{\textrm{P}}$ and $x^\mu_{\textrm{Q}}$ is given by 
\begin{equation}
L=\int_{x^\mu_{\textrm{P}}}^{x^\mu_{\textrm{Q}}} d\sigma \sqrt{g_{\mu\nu}s^\mu(\sigma) s^\nu(\sigma)}\,,
\end{equation}
where $s^\mu(\sigma)$ is the tangent vector to a space-like curve that connects the segments of the rod from the mirror at $x^\mu_{\textrm{P}}$ to the mirror at $x^\mu_{\textrm{Q}}$ chosen such that it lies inside the spatial slice defined by the rod's rest frame \cite{Raetzelfreq:2018}. 
To quantize the light field in the vertical and horizontal cavities, we describe the electromagnetic field using a massless scalar field $\psi$ obeying the Klein-Gordon equation
\begin{align} \label{eq:klein_gordon}
\square \psi = \frac{1}{\sqrt{-g}}\partial_\mu \left(\sqrt{-g}\,g^{\mu \nu} \partial_\nu \psi \right)\,,
\end{align}
where $g := \text{det}(\boldsymbol{g})$ is the metric \cite{birrell}. This is a good approximation when considering a single polarization mode of the electromagnetic field \cite{PhysRevD.88.064028}. The general case must be treated with a spin-1 quantum field. The field can be written as a sum of the mode solutions $\phi_{k}$ of the Klein-Gordon equation,
\begin{align} \label{eq:klein_gordon:full:solution}
\psi = \sum_k\left[\alpha_k\,\phi_{k}+\alpha^*_k\,\phi_{k}^*\right].
\end{align}
The mode numbers $k$ are discrete because the solutions must vanish at the mirrors.  $\alpha_k$ are the time-independent Fourier coefficients, which are promoted to operators upon quantization \cite{birrell}. In the following we will quantize the field in the horizontal and vertical cavities using the metric approximation derived in the previous section. 

\subsection{Horizontal cavities}
A horizontal cavity is placed at $x = \bar{r}_0$ with the length of the cavity oriented in the $z$-direction as shown in Figure~\ref{fig:coordinates}. The cavity mirrors are located at $z=z_{\textrm{l}}=-l/2$ and $z=z_{\textrm{r}}=l/2$, with $l:=z_{\textrm{r}}-z_{\textrm{l}}>0$ (see also Figure~\ref{fig:cross}). The setup has rotational symmetry in the $y$-$z$-plane. A convenient choice of coordinates is $(x^0,z)$, where $\bar{r}(z):=(\bar{r}_0^2+z^2)^{1/2}$. The metric components (\ref{eq:schwarzschildapp}) can be specified for the horizontal cavity and expanded up to second order in $z/\bar{r}_0$ and $r_{\textrm{S}}/\bar{r}_0$ and to first order in their product
\begin{eqnarray}\label{eq:hormet}
	\boldsymbol{g}_{\textrm{h}}&=&f_\rm{RS}(\bar{r}_0)\rm{diag}\left(-1,\Sigma(\bar{r}_0)\right)\,,
\end{eqnarray}
where
\begin{equation*}
	f_\rm{RS}(\bar{r}):=\left(1-\frac{r_{\textrm{S}}}{\bar{r}}+\frac{r_{\textrm{S}}^2}{2\bar{r}^2}\right)\,,
\end{equation*}
is the redshift function and 
\begin{equation*}
	\Sigma(\bar{r}):=\left(1+\frac{2r_{\textrm{S}}}{\bar{r}}+\frac{15r_{\textrm{S}}^2}{8\bar{r}^2}\right)\,,
\end{equation*}
the spatial scale function. The proper length of the horizontal rod is given by
\begin{align}
L_{\textrm{h}} \approx \int_{-l/2}^{l/2}dz\,f_\rm{RS}^{1/2}(\bar{r}_0)\Sigma^{1/2}(\bar{r}_0) = f_\rm{RS}^{1/2}(\bar{r}_0)\Sigma^{1/2}(\bar{r}_0)\,l
\end{align}
where we used the four-vector $s^\mu(\sigma)=(0,0,0,1)$. The position of the left and right mirrors are given as a function of the proper length through 
\begin{align}
    z_{\textrm{l}} \approx -f_\rm{RS}^{-1/2}(\bar{r}_0)\Sigma^{-1/2}(\bar{r}_0)\frac{L_\textrm{h}}{2}\,\nonumber\\
	z_{\textrm{r}} \approx f_\rm{RS}^{-1/2}(\bar{r}_0)\Sigma^{-1/2}(\bar{r}_0)\frac{L_\textrm{h}}{2}.
\end{align}
The solutions to the Klein-Gordon equation are conformally-invariant in $1+1$ dimensions \cite{birrell}.
Therefore, it is convenient to make a conformal transformation to coordinates $(x^0,\tilde{z})$ where $\tilde{z}=\Sigma^{1/2}(\bar{r}_0)z$. This can be done easily since the metric $\boldsymbol{g}_{\textrm{h}}$ does not depend on the variable $z$. 
In conformal coordinates, the metric
\begin{eqnarray}\label{eq:hkg}
	\tilde{\boldsymbol{g}}_{\textrm{h}}&=&f_\rm{RS}(\bar{r}_0)\rm{diag}\left(-1,1\right),
\end{eqnarray}
and the proper length $L_{\textrm{h}} \approx f_\rm{RS}^{1/2}(\bar{r}_0)\,(\tilde z_{\textrm{r}}-\tilde{z}_{\textrm{l}})$ take a simple form, where we have defined $\tilde{z}_{\textrm{l}}:=\Sigma^{1/2}(\bar{r}_0)z_{\textrm{l}}$ and $\tilde{z}_{\textrm{r}}:=\Sigma^{1/2}(\bar{r}_0)z_{\textrm{r}}$.

The solutions to the Klein-Gordon equation (\ref{eq:klein_gordon}) with this background metric must vanish at the mirrors. Therefore, we solve the equation and impose Dirichlet boundary conditions $\psi_{\textrm{h}} (x^0,\tilde z_{\textrm{l}}) = \psi_{\textrm{h}} (x^0,\tilde z_{\textrm{r}}) = 0$. The mode solutions $\phi_{\textrm{h},k}$ are
\begin{align}\label{eq:hsol}
\phi_{\textrm{h},k}(x^0,\tilde z) := \frac{e^{-i\Omega_{\textrm{h},k} t}}{\sqrt{\pi k}}\sin \left( \frac{\Omega_{\textrm{h},k}}{c}(\tilde z-\tilde{z}_{\textrm{l}})\right),
\end{align}
where $k$ is a positive integer. The frequencies $\Omega_{\textrm{h},k}$ are given by
\begin{align}\label{horizontal:fake:frequency}
\Omega_{\textrm{h},k} = \frac{c\pi k}{\tilde z_{\textrm{r}}-\tilde{z}_{\textrm{l}}} = \frac{c\pi k}{L_{\textrm{h}}}f_\rm{RS}^{1/2}(\bar{r}_0).
\end{align}
The mode solutions are normalized with respect to the (horizontal) Klein-Gordon inner product 
\begin{align}
(\phi_{\textrm{h},k},\phi_{\textrm{h},k'})=&\,\delta_{kk'}\nonumber\\
=&\,i \int_{\tilde{z}_{\textrm{l}}}^{\tilde z_{\textrm{r}}} d\tilde z f^{-1/2}(\bar{r}_0)(\phi_{\textrm{h},k}^{*}\partial_0\phi_{\textrm{h},k'} - {\phi}_{\textrm{h},k}\partial_0\phi_{\textrm{h},k'}^*)\,.
\end{align} 
The field $\psi_{\textrm{h}}$ is quantized in the $\phi_{\textrm{h},k}$ basis by associating annihilation and creation operators $\hat{a}_{\textrm{h},k}$ and $\hat{a}_{\textrm{h},k}^\dag$ to the positive and negative solutions $\phi_{\textrm{h},k}$ and $\phi_{\textrm{h},k}^*$ respectively. The quantized field in the horizontal cavity is
\begin{equation} \label{eq:horizontal_free_field}
\hat{\psi}_{\textrm{h}}(x^0,\tilde z) = \sum_k \left\{ \hat{a}_{\textrm{h},k}\,\phi_{\textrm{h},k}(x^0,\tilde z)  + \hat{a}_{\textrm{h},k}^\dag\,\phi_{\textrm{h},k}^*(x^0,\tilde z) \right\}\,.
\end{equation}
The creation and annihilation operators satisfy the canonical commutation relations $[\hat{a}_{\textrm{h},k},\hat{a}_{h,k'}^\dag] = \delta_{k,k'}$, while all other commutators vanish. 

\subsection{Vertical cavities}
\begin{figure}[ht!]
\includegraphics[width=\linewidth]{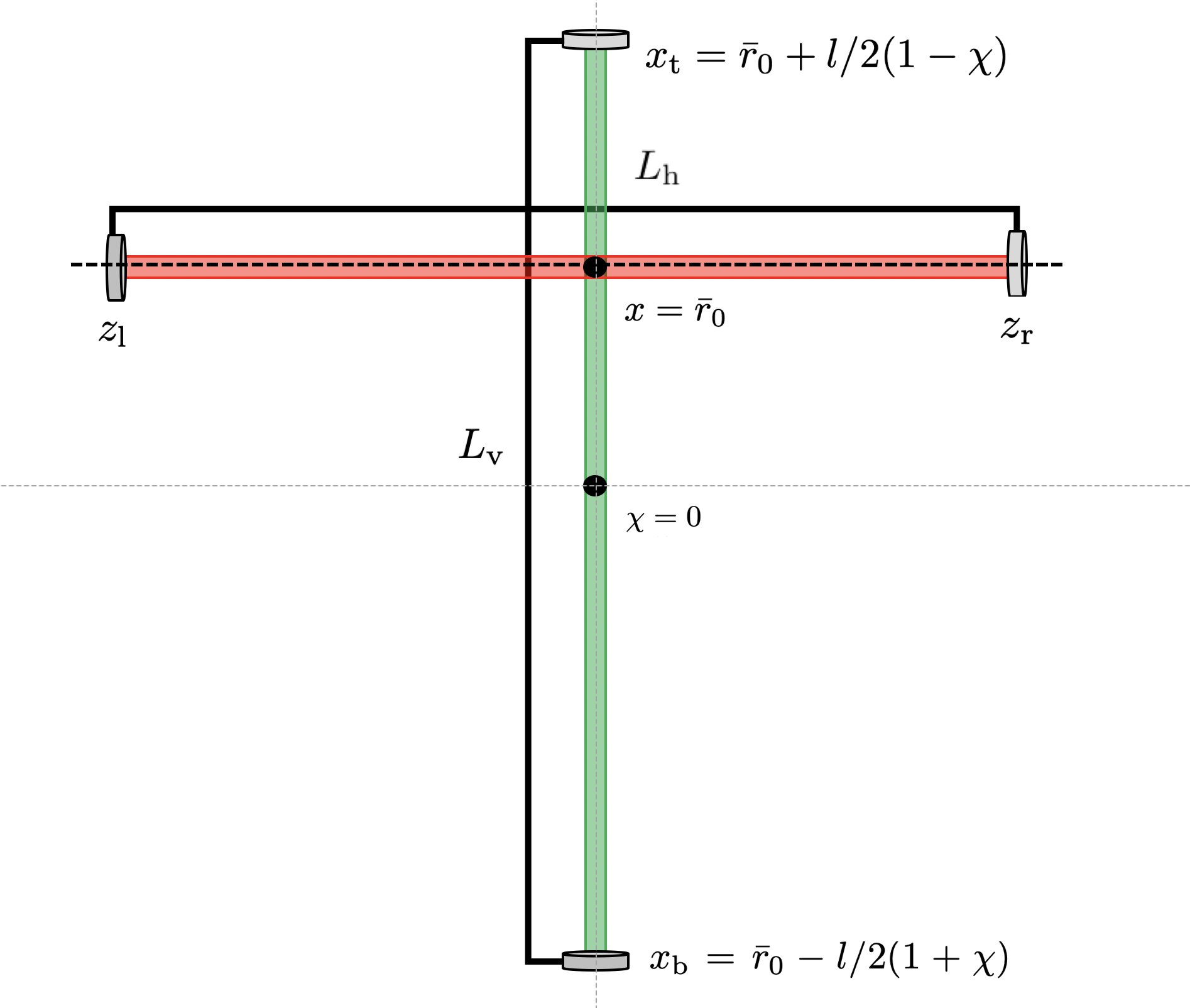}
\caption{The horizontal and vertical cavities intersect at $x=\bar{r}_0$. This figure shows where the mirrors are located.}
\label{fig:cross}
\end{figure}

A vertical cavity with two mirrors at its ends is placed along the $x$-axis and intersects the horizontal cavity at $x=\bar{r}_0$. We assume that the top mirror is located $x=x_{\textrm{t}}=\bar{r}_0+l(1-\chi)/2$, while the bottom mirror is at $x=x_{\textrm{b}}=\bar{r}_0-l(1+\chi)/2$, with $-1\leq\chi\leq 1$. For example, $\chi=-1$ corresponds to the cavities intersecting at the bottom point of the vertical cavity, while for $\chi=1$ the intersection is at the top. The ``cross configuartion'' corresponds to $\chi=0$. The point of intersection between the cavities will be used as a reference to define the frequencies measured in the laboratory frame in section \ref{fig:cross}.
Now it is convenient to choose coordinates $(x^0,x')$, where the origin is at the intersection with the horizontal cavity $x'=x-\bar{r}_0$ and $x_b-\bar{r}_0\le x' \le x_t - \bar{r}_0$. If the cavity's proper length is such that $L_{\textrm{v}}/\bar{r}_0\sim r_{\textrm{S}}/\bar{r}_0$, then in the new coordinates, $x'/\bar{r}_0\ll 1$ and $x'/\bar{r}_0\sim r_{\textrm{S}}/\bar{r}_0$. The effective $1+1$-dimensional metric inside the vertical cavity given up to second order in $r_{\textrm{S}}/\bar{r}_0$ and $x'/\bar{r}_0$ and up to first order in their product is 
\begin{equation}\label{eq:vertmet}
	\boldsymbol{g}_{\textrm{v}}= f_\rm{RS}(\bar{r}_0)\left(1+\frac{r_{\textrm{S}}x'}{\bar{r}_0^2}\right)\rm{diag}\left(-1,\Sigma(\bar{r}_0)\left(1-\frac{2r_{\textrm{S}}x'}{\bar{r}_0^2}\right)\right).
\end{equation}
The proper length of the vertical rod is given by 
\begin{align}
\nonumber L_{\textrm{v}}&=\int^{l(1-\chi)/2}_{-l(1+\chi)/2} dx'\,(\boldsymbol{g}_{\textrm{v},{11}}(x))^{1/2}\\&=f_\rm{RS}^{1/2}(\bar{r}_0)\Sigma^{1/2}(\bar{r}_0)\,l\,\left(1+\frac{r_{\textrm{S}}\,l}{4\bar{r}_0^2}\chi\right)
\end{align}
where we used that $s^\mu(\sigma)=(0,1,0,0)$, $x^1_{\textrm{Q}}=x_{\textrm{b}}$ and $x^1_{\textrm{P}}=x_{\textrm{t}}$. The positions of the bottom and top mirrors with respect of the intersection point are given in terms of the proper length by the following expressions
\begin{align}\label{eq:xprime}
	x'_{\textrm{b}} = & -f^{-1/2}_\rm{RS}(\bar{r}_0)\Sigma^{-1/2}(\bar{r}_0) \frac{1+\chi}{2}\left(1-\frac{r_{\textrm{S}}L_{\textrm{v}}}{4\bar{r}_0^2}\chi\right)L_{\textrm{v}}\nonumber\\
	x'_{\textrm{t}} = & f^{-1/2}_\rm{RS}(\bar{r}_0)\Sigma^{-1/2}(\bar{r}_0) \frac{1-\chi}{2}\left(1-\frac{r_{\textrm{S}}L_{\textrm{v}}}{4\bar{r}_0^2}\chi\right)L_{\textrm{v}}.
\end{align}
A conformal transformation to coordinates  $(x^0,\tilde{x})$ is made in order to solve the Klein-Gordon equation
\begin{eqnarray}\label{eq:xtildetrans}
	\nonumber \tilde{x}&=&\Sigma^{1/2}(\bar{r}_0)\int_0^{x'} dx''\,\left(1-\frac{2r_{\textrm{S}}x''}{\bar{r}_0^2}\right)^{1/2}\\
	 &\approx &\Sigma^{1/2}(\bar{r}_0)x'\left(1-\frac{r_{\textrm{S}}x'}{2\bar{r}_0^2}\right)\,,
\end{eqnarray}
which leads to $x'=\Sigma(\bar{r}_0)^{-1/2}\tilde{x}(1+r_{\textrm{S}}\tilde{x}/2\bar{r}_0^2)$. In conformal coordinates, the vertical metric to second order in $r_{\textrm{S}}/\bar{r}_0$ and $\tilde{x}/\bar{r}_0$, and to first order in their product, is given by
\begin{equation}\label{eq:vkg}
	\tilde{\boldsymbol{g}}_{\textrm{v}} = f_\rm{RS}(\bar{r}_0)\left(1+\frac{r_{\textrm{S}}\tilde{x}}{\bar{r}_0^2}\right)\rm{diag}\left(-1,1\right)\,.
\end{equation}
The Dirichlet boundary conditions are $\psi_{\textrm{v}} (x^0,\tilde{x}_b) = \psi_{\textrm{v}} (x^0,\tilde{x}_{\textrm{t}}) = 0$, where
\begin{align}\label{eq:xtilde:top:bottom}
	\tilde{x}_{\textrm{b}} = & -f^{-1/2}_\rm{RS}(\bar{r}_0) \frac{1+\chi}{2}\left(1+\frac{r_{\textrm{S}}L_{\textrm{v}}}{4\bar{r}_0^2}\right)L_{\textrm{v}}\nonumber\\
	\tilde{x}_{\textrm{t}} = & f^{-1/2}_\rm{RS}(\bar{r}_0) \frac{1-\chi}{2}\left(1-\frac{r_{\textrm{S}}L_{\textrm{v}}}{4\bar{r}_0^2}\right)L_{\textrm{v}}.
\end{align}
are the positions of the mirrors in the conformal coordinates obtained through equations \eqref{eq:xprime} and \eqref{eq:xtildetrans}. The mode solutions yield
\begin{align}\label{mode:solution:vertical}
\phi_{\textrm{v},k}(x^0,\tilde{x}) := \frac{e^{-i\Omega_{\textrm{v},k} t}}{\sqrt{\pi k}}\sin \left( \frac{\Omega_{\textrm{v},k}}{c}\tilde{x}\right),
\end{align}
where the frequencies $\Omega_{\textrm{v},k} := \frac{c\pi k}{\tilde{x}_{\textrm{t}}-\tilde{x}_{\textrm{b}}}$ are given in terms of  the proper length $L_{\textrm{v}}$, and read
\begin{equation}\label{vertical:fake:frequency}
	\Omega_{\textrm{v},k} \approx  \frac{c\pi k}{L_{\textrm{v}}}f^{1/2}_\rm{RS}(\bar{r}_0) 
	\left(1 - \frac{r_{\textrm{S}}L_{\textrm{v}}}{4\bar{r}_0^2}\chi\right).
\end{equation}
To quantize the field, we introduce creation and annihilation operators $\hat{a}_{\textrm{v},k}^\dag,\hat{a}_{\textrm{v},k}$ that satisfy the canonical commutation relations $[\hat{a}_{\textrm{v},k},\hat{a}_{\textrm{v},k'}^\dag] = \delta_{k,k'}$, while all other commutators vanish. The quantized field $\hat{\psi}_{\textrm{v}}$ in the vertical cavity is given by
\begin{equation} \label{eq:free_field}
\hat{\psi}_{\textrm{v}}(x^0,\tilde{x}) = \sum_k  \left\{ \hat{a}_{\textrm{v},k}\,\phi_{\textrm{v},k}(x^0,\tilde{x})  + \hat{a}_{\textrm{v},k}^\dag\,\phi_{\textrm{v},k}^*(x^0,\tilde{x}) \right\}.
\end{equation}

\section{\label{sec:clocks}Quantum clocks in Schwarzschild spacetime}

A single mode of the quantized cavity field of the previous sections can be used to measure time since it oscillates at a fixed frequency.
  In this section, we investigate the properties of these quantized light clocks in a Schwarzschild spacetime. In particular,
 quantum and classical states will be used to study the effects of spacetime on the precision of the horizontal and vertical light clocks. While coherent states approximate well classical states of light when the number of photons is high, squeezed states are commonly used in quantum sensing since they are known to lead to high precision \cite{Giovannetti2006quant}. Recently, squeezed states have been used in the Laser Interferometric Gravitational Observatory (LIGO) and the Virgo Collaboration to improve the detection of gravitational waves \cite{Tse:2019,Schnabel:2016gdi, Acernese:2019}. The quantized light model includes both quantum and relativistic effects, enabling the study of time at the interplay of these theories.

\subsection{Local frequencies}
To define the frequencies that are measured in the experiment, it is necessary to introduce a laboratory reference frame. The mode solutions \eqref{eq:hsol} and \eqref{mode:solution:vertical} are given in terms of the coordinate time $t$, which is the time measured by an ideal point-like clock at infinity. The frequencies $\Omega_{\textrm{h},k}$ and $\Omega_{\textrm{v},k}$, obtained in \eqref{horizontal:fake:frequency} and \eqref{vertical:fake:frequency}, are defined with respect to this time. The horizontal cavity can be used as a laboratory reference clock by defining the frequencies with respect to the point $\bar{r}_0$ where the horizontal cavity intersects with the vertical one.

To introduce the laboratory reference frame, it is necessary to provide a prescription for the positions of the mirrors. Initially, two identical cavities with the same length $L$ are placed horizontally at $\bar{r}_0$. This will be the reference length. One of the cavities is subsequently rotated and placed vertically along the $x$ axis as shown in the Figure (\ref{fig:cross}). Assuming that the bars that hold the mirrors are infinitely rigid, yields the condition
\begin{align}
L\equiv L_{\textrm{h}}=L_{\textrm{v}}\,.
\end{align}

In special relativity time depends on the state of motion of the observer. A meaningful notion of time is constructed by postulating that the proper time $\tau:=\int_{\textrm{P}}^{\textrm{Q}} \sqrt{-ds^2}/c$ measures the time of a point-like observer that moves along the world-line between points P and Q , where $ds^2$ is the line element. The proper time of an observer static at a fixed radial position $r$ in Schwarzschild spacetime is given by $\tau(r)= f(r)^{1/2} t$ with respect to the coordinate time $t$. The time dilation $f_\rm{RS}^{1/2}(\bar{r}_0):=\tau_{\textrm{0}}/t$ is the redshift between the time measured by a point-like clock at $\bar{r}_0$ and one at infinity. We also make reference to the proper time $\tau_0=f_\rm{RS}^{1/2}(\bar{r}_0)\,t$ measured by an ideal point-like clock at $\bar{r}_0$ solely for conceptual matters. The time $\tau_0$ is a common book-keeping reference useful in defining the frequencies. This is not trivial since in the vertical cavity, the proper time flows differently at each point. The frequencies in the horizontal and vertical cavities measured in the laboratory frame are defined with respect to this unique common point through the relations $\omega_{\textrm{h},k}\tau_0=\Omega_{\textrm{h},k}\,t$ and $\omega_{\textrm{v},k}\,\tau_0=\Omega_{\textrm{v},k}\,t$. This yields,

\begin{align}\label{eq:omega:h:v}
\omega_{\textrm{h},k}(\bar{r}_0)&=\frac{c\pi k}{L}\nonumber\\
 \omega_{\varv,k}(\bar{r}_0)&=\omega_{\textrm{h},k}\left(1 - \frac{r_{\textrm{S}}L}{4\bar{r}_0^2}\chi\right).
\end{align}
In the case that $\bar{r}_0\rightarrow\infty$, the frequency $\omega_{\infty,k}:=\frac{c\pi k}{L}$ corresponds to the frequency measured in flat spacetime with respect to the coordinate time $t$.  In this limit, the horizontal and vertical frequencies coincide $\omega_{\textrm{h},k}(\infty)=\omega_{\textrm{v},k}(\infty)=\omega_{\infty,k}$ in agreement with the literature \cite{Pound:1959grav}. 

 From equation (\ref{eq:omega:h:v}), we also find that the frequency of the horizontal clock is equal to the frequency of a clock of length $L$ at infinity.
In contrast, the frequency of the vertical clock is modified depending on the parameter $\chi$ that parameterizes the radial position of the vertical cavity with respect to the horizontal cavity.
The relative frequency difference is
\begin{align}\label{relative:frequency:difference}
\frac{\delta\omega_{k}}{\omega_{\textrm{h},k}}:=\frac{\omega_{\varv,k}- \omega_{\textrm{h},k}}{\omega_{\textrm{h},k}}&=-\frac{r_{\textrm{S}}L}{4\bar{r}_0^2}\chi=-\frac{GML}{2c^2\bar{r}_0^2}\chi
\end{align}
This correction is proportional to $1/c^2$, therefore, relativistic.
 $\chi=-1$ corresponds to the case when the clocks intersect at the lower vertical mirror, $\chi=1$ when they intersect at the top mirror and, $\chi=0$ when they intersect at the center of the vertical clock. When the clocks intersect at the center, there is no correction. The correction changes sign depending on the intersection being towards the top or bottom mirror.  
 In general, we can interpret the dependence on $\chi$ as the result of gravitational red shift/time dilation: we have found that the light clock ticks with the proper time defined by its center. If one compares the ticking of the light clock with the proper time at any other position, one obtains a gravitational time dilation.



\subsection{Gaussian states and Quantum Metrology techniques}
The precision of quantum light clocks can be estimated using quantum metrology techniques which we will introduce in the following.
 Quantum metrology provides strategies to estimate physical parameters such as time, mass and field strengths. These strategies enable precisions that are higher in comparison to those reached by classical methods \cite{Giovannetti2006quant}. Quantum metrology strategies include finding optimal quantum input states and measurements. The input state is given by the density matrix $\hat{\rho}(0)$.  We assume that a unitary channel $\hat{U}(\lambda)$ acts on the state such that the final state $\hat{\rho}(\lambda):=\hat{U}(\lambda)\hat{\rho}(0)\hat{U}^\dag(\lambda)$ encodes the parameter $\lambda$ that will be estimated. In this work, the unitary channel is the time evolution of the modes. The precision on the measurement of $\lambda$ is bounded from below by the quantum Cram\'er-Rao bound \cite{helstrom1976quantum,holevo1982probabilistic,Braunstein:1994} ,
\begin{equation}
\Delta \lambda \ge \frac{1}{\sqrt{\mathcal{M}\,H(\lambda)}}\,
\end{equation}
where \cite{hayashi2006quantum}
\begin{equation}
H(\lambda):=\underset{d\lambda\rightarrow0}{\textrm{lim}}8\frac{1-\sqrt{\mathcal{F}(\hat{\rho}(\lambda),\hat{\rho}(\lambda+d\lambda))}}{d\lambda^2},
\end{equation}
is the Quantum Fisher Information (QFI) and $\mathcal{M}$ the number measurements. The function  $\mathcal{F}(\hat{\rho},\hat{\rho}'):=[\textrm{Tr}(\sqrt{\sqrt{\hat{\rho}}\,\hat{\rho}'\,\sqrt{\hat{\rho}}})]^{2}$ is the fidelity between states $\hat{\rho}$ and $\hat{\rho}'$. Given an input state, the Cram\'er-Rao bound optimizes over all possible quantum measurements. 

 The optimal measurement is usually hard to find mathematically. If found, then it might be hard to implement in the experiment. Therefore, it is convenient to study the precision obtained using measurements that are commonly preformed in the laboratory and compare it to the optimal precision bound. In the case of light modes, homodyne and heterodyne measurements are easy to implement in the laboratory using photo-counters and linear-optical elements such as mirrors and beam-splitters. The analysis of clock precision using quantum light modes will be restricted to Gaussian states because in this case homodyne and heterodyne measurements yield precisions that approximate closely the optimized precision bound. Gaussian states not only have practical advantages in the laboratory, but also mathematical ones. Gaussian states are uniquely characterized by their first and second moments. Higher moments vanish in this case. The state of the field can be described using the field moments instead of the density matrix simplifying most calculations. 

 To define the field moments, the 
 canonical operators $\hat{a}_n,\hat{a}^\dag_n$ of $N$ modes are collected in the vector $\hat{\mathbb{X}}:=(\hat{a}_1,...,\hat{a}_N,\hat{a}^\dag_1,...,\hat{a}^\dag_N)^{\textrm{Tp}}$. The first moments are given by the vector with components $d_n:=\langle\hat{X}_n\rangle$ and the $2N\times2N$ covariance matrix, defined through its elements $\Gamma_{nm}:=\langle\{\hat{X}_n,\hat{X}_m^\dag\}\rangle-2\langle\hat{X}_n\rangle\langle\hat{X}_m^\dag\rangle$, contains all second moments of the field. The expectation values are taken with respect to the initial state, since we work in the Heisenberg picture. The function $\{\cdot,\cdot\}$ is the anticommutator. 

 Gaussian states remain Gaussian under unitary transformations $\hat{U}(\lambda)$ that are quadratic in the field operators. The transformation is \textit{linear} since  $\hat{U}^\dag(\lambda)\,\hat{\mathbb{X}}\,\hat{U}(\lambda)=\boldsymbol{S}(\lambda)\,\hat{\mathbb{X}}$.  The matrix $\boldsymbol{S}(\lambda)$ is a $2N\times2N$\textit{symplectic} matrix that is an equivalent representation of the operator $\hat{U}(\lambda)$. If the unitary transformation is induced by a time independent quadratic Hamiltonian $\hat{U}(\lambda)=\exp[-i \hat{H}\,\lambda/\hbar]$, then $\boldsymbol{S}(\lambda)=\exp[\boldsymbol{\Omega}\,\boldsymbol{H}\,\lambda]$. The Hamiltonian matrix $\boldsymbol{H}$ is defined by the relation
 $\hat{H}:=\hbar\hat{\mathbb{X}}^\dag\,\boldsymbol{H}\,\hat{\mathbb{X}}/2$, and $\boldsymbol{\Omega}:=-i\,\textrm{diag}(1,...,1,-1,...,-1)$ is known as the \textit{symplectic form} \cite{vsafranek2016optimal}. 

When the transformation $\hat{\rho}(\lambda):=\hat{U}(\lambda)\hat{\rho}(0)\hat{U}^\dag(\lambda)$ corresponds to a linear transformation on a Gaussian state, it can be mapped to $\boldsymbol{\Gamma}(\lambda)=\boldsymbol{S}(\lambda)\,\boldsymbol{\Gamma}(0)\,\boldsymbol{S}^\dag(\lambda)$and $\boldsymbol{d}(\lambda)=\boldsymbol{S}(\lambda)\,\boldsymbol{d}(0)$. In this case, the quantum Fisher information takes a simple form \cite{dom}
\begin{align}
H(\lambda) =& \frac{1}{4}\text{Tr}\left[\left(\boldsymbol{\Gamma}(\lambda)^{-1}
\dot{\boldsymbol{\Gamma}}(\lambda)\right)^2\right]+ 2\dot{\boldsymbol{d}}^\dag(\lambda) \boldsymbol{\Gamma}^{-1}(\lambda) \dot{\boldsymbol{d}}(\lambda).
\end{align}
Here the dot represents the derivative with respect to $\lambda$. To estimate the precision with which quantized light clocks measure proper time using the quantum Fisher information, the covariance matrix $\boldsymbol{\Gamma}(\lambda=\tau_0)$ and the displacement vector $\boldsymbol{d}(\lambda=\tau_0)$ for an input clock state $\boldsymbol{\Gamma}(0)$, $\boldsymbol{d}(0)$ will be computed in the next section.

 \subsection{Input clock states}

To prepare the input clock state, the mode $k$ is squeezed and displaced using the operators $\hat{S}_k(\mathrm{r})=\text{exp}[\frac{\mathrm{r}}{2}(\hat{a}_k^2 - \hat{a}_k^{\dag 2})]$ and $\hat{D}_k(\alpha)=\text{exp}[\alpha\hat{a}_k^{\dag} - \alpha^*\hat{a}_k]$, respectively. The resulting state is called a single mode squeezed coherent (SMSC) state   $|\alpha,\mathrm{r}\rangle:=\hat{D}_k(\alpha)\hat{S}_k(\mathrm{r}) |0_k\rangle$. The state $|0_k\rangle$ is the vacuum state, defined by the relation $\hat{a}_k|0_k\rangle=0$. The real parameter $\mathrm{r} \in \mathbb{R}$ is called squeezing parameter, while the complex $\alpha=|\alpha|\exp[i\phi] \in \mathbb{C}$ is called displacement. The physical interpretation of these operations can be obtained using a phase-space representation through the Wigner function \cite{Weedbrook}. The squeezing operator reduces the variance along one axis, at the expense of the variance of the orthogonal axis (for example of the position at the expense of the momentum), while the displacement shifts rigidly the distribution. The number of photons in a squeezed coherent state is given through $\alpha$ and $\mathrm{r}$ as $N_p=\sinh^2(\mathrm{r}) + |\alpha|^2e^{2\mathrm{r}}$.

The $2\times2$ covariance matrix of the state $|\alpha,\mathrm{r}\rangle$ is
\begin{align}
\boldsymbol{\Gamma}^{\text{SMSC}}_{k}(0)=
\begin{pmatrix}
\cosh \mathrm{(2r)} & -\sinh \mathrm{(2r)} \\
-\sinh \mathrm{(2r)} & \cosh \mathrm{(2r)}
\end{pmatrix}.
\end{align}
and the displacement vector is $\boldsymbol{d} = (\alpha, \alpha^*)^{\textrm{Tp}}$. The clock is stationary, therefore, the state of the field mode only undergoes the free evolution  \begin{align}
\boldsymbol{\Gamma}^{\text{SMSC}}_{k}(t)=\boldsymbol{S}_0(t)\boldsymbol{\Gamma}^{\text{SMSC}}_{k}(0)\boldsymbol{S}_0^\dag(t),
\end{align}
where the free evolution symplectic matrix is given by
\begin{align}\label{free:time:evolution:matrix}
\boldsymbol{S}_0(t)=
\begin{pmatrix}
e^{-i\psi_{k}(t)}&0 \\
0 & e^{i\psi_k(t)}
\end{pmatrix},
\end{align}
The phase is given by $\psi_{k}(t)=\Omega_{\textrm{h,}k}t$ or $\psi_{k}(t)=\Omega_{\textrm{v},k}t$ depending on the clock being either horizontal or vertical. The state evolution (\eqref{free:time:evolution:matrix}) is a unitary channel that encodes the time on the input state $\boldsymbol{\Gamma}^{\text{SMSC}}_{k}(0)$. As a consequence of this process, the time parameter $t$ becomes encoded in the state $\boldsymbol{\Gamma}^{\text{SMSC}}_{k}(t)$.

\subsection{Precision for proper time measurements}
Once the input state $|\alpha,\mathrm{r}\rangle$ has been specified and the final state obtained after considering free evolution \eqref{free:time:evolution:matrix}, the quantum Fisher information can be computed to find the bound on the precision of the light clocks,
\begin{align}\label{eq:qfi_single_mode_general}
H^{\text{SMSC}}(\tau_0) = & \,2 \left[\sinh^2(2\mathrm{r}) + 2|\alpha|^2(\cosh(2\mathrm{r})\right.\,,\nonumber\\
&\left.-\cos(2\phi)\sinh(2\mathrm{r}))\right]\dot{\psi}^2_k(\tau_0) 
\end{align}
where the dot denotes derivative with respect to the proper time $\tau_0$ at the point where the horizontal and vertical clocks intersect. 
Optimization of the expression above with respect to the phase $\phi$ of the displacement can be done once the sign of $\mathrm{r}$ has been fixed. Choosing, without loss of generality, that $\mathrm{r}\geq0$, we therefore have that the largest QFI is obtained for $\phi=(n+1/2)\,\pi$, and it reads
\begin{align}\label{eq:qfi_single_mode_general:optimal}
H^{\text{SMSC}}(\tau_0) = & \,2 \left[\sinh^2(2\mathrm{r}) + 2|\alpha|^2e^{2\,\mathrm{r}}\right]\dot{\psi}^2_k(\tau_0)\,. 
\end{align}
The phase $\psi_k(\tau_0)$ for horizontal and vertical clocks is given by $\psi_{\textrm{h},k}(\tau_0):=\omega_{\textrm{h},k}\,\tau_0$ and $\psi_{\textrm{v},k}(\tau_0):=\omega_{\textrm{v},k}\,\tau_0$. Therefore, the bound on the absolute error in estimating the proper time $\tau_0$ is given in the two cases by 
\begin{align}\label{final:absolute:precision:bounds}
\Delta_{\textrm{h},k}(\tau_0) &= \frac{1}{\sqrt{2\,\mathcal{M}}}\frac{1}{\sqrt{\sinh^2(2\mathrm{r}) + 2|\alpha|^2e^{2\mathrm{r}}} }\frac{L}{\pi\,c\,k}\nonumber\\
\Delta_{\textrm{v},k}(\tau_0) &=\Delta_{\textrm{h},k}(\tau_0) \left(1 + \frac{r_{\textrm{S}}L}{4\bar{r}_0^2}\chi\right)
\end{align}
to first order in $r_{\textrm{S}}L/\bar{r}_0^2$.
The lower bound approaches infinity as the cavity length approaches zero. However, in this limit the frequencies diverge. It is also interesting to note that the precisions \eqref{final:absolute:precision:bounds} are independent of time.

The bound to the relative error $\delta(\tau_0) := |\Delta \tau_0|/|\tau_0|$, is given by
\begin{align}\label{final:relative:precision:bounds}
\delta_{\textrm{h},k}(\tau_0) &=\frac{\Delta_{\textrm{h},k}(\tau_0)}{\tau_0}\,,\nonumber\\
\delta_{\textrm{v},k}(\tau_0) &=\delta_{\textrm{h},k}(\tau_0)\left(1 + \frac{r_{\textrm{S}}L}{4\bar{r}_0^2}\chi\right).
\end{align}
The total number of field mode oscillations $\frac{ck\pi}{L}$ is the number of clock ticks. The number of clicks and the precision of a horizontal clock at $\bar{r}_0$ is the same as of clocks at infinity. However, the number of clicks for the precision of the vertical clock is slightly larger due to gravitational time dilation.
Increasing the time increases the relative precision.

\subsection{Comparison between classical and quantum light clocks}
 In the case that $|\alpha|$ is large, coherent states behave as classical states. Therefore, a classical light clock corresponds to an input state where $|\alpha|\gg 1$ and $\mathrm{r}=0$. On the other hand, a squeezed vacuum state where $|\alpha|=0$ is a state with strong non-classical properties. 
 To compare the performance of classical and quantum clocks, it is convenient to fix the average number of excitations $N_p$ in the clock state. For classical clocks $N_p:=|\alpha|^2$ and for the squeezed vacuum quantum clock $N_p:=\sinh^2\mathrm{r}$. The precision bounds \eqref{final:absolute:precision:bounds} for horizontal and vertical classical clocks are
\begin{align}\label{final:absolute:precision:bounds:coherent state}
\Delta_{\textrm{h},k}(\tau_0) &= \frac{1}{2\sqrt{\mathcal{M}}}\frac{1}{\sqrt{N_p}}\frac{L}{\pi\,c\,k}\nonumber\\
\Delta_{\textrm{v},k}(\tau_0) &=\Delta_{\textrm{h},k}(\tau_0) \left(1 + \frac{r_{\textrm{S}}L}{4\bar{r}_0^2}\chi\right).
\end{align}
And for quantum clocks with squeezed vacuum input states
\begin{align}
\Delta_{\textrm{h},k}(\tau_0) &= \frac{1}{2\sqrt{\mathcal{M}}}\frac{1}{\sqrt{N_p(N_p+1)}}\frac{L}{\pi\,c\,k}\nonumber\\
\Delta_{\textrm{v},k}(\tau_0) &=\Delta_{\textrm{h},k}(\tau_0) \left(1 + \frac{r_{\textrm{S}}L}{4\bar{r}_0^2}\chi\right).
\end{align}
For the same number of resources (excitations in the input state), quantum light clocks estimate proper time with higher precision. The quantum clock has a Heisenberg scaling $\approx 1/N_p$, while the scaling for the classical clock is $1/\sqrt{N_p}$.



\subsection{Estimating the Schwarzschild radius}

Depending on the parameter $\chi$, the two differently oriented quantized light clocks tick with different rates. Therefore, a local clock comparison can be used to estimate the acceleration, and thus, the Schwarzschild radius of the spacetime. As the two clocks overlap, the clock comparison may be realized, for example, by bringing light from the two clock modes into interference at a beam splitter. 

Formally, we obtain the strength of the clock rate deviation induced by $r_\mathrm{S}$ through the derivative of the phase $\psi_{\textrm{v},k}=\omega_{\mathrm{v},k}t$ with respect to $r_\mathrm{S}$. The corresponding QFI becomes
\begin{align}
H^{\text{SMSC}}(r_S) = & \,2 \left[\sinh^2(2\mathrm{r}) + 2|\alpha|^2e^{2\,\mathrm{r}}\right]\left(\frac{d \psi_{\textrm{v},k}}{dr_{\textrm{S}}}\right)^2\,,
\end{align}
and accordingly 
\begin{align}
\Delta_{k}(r_{\textrm{S}}) &= \frac{1}{\sqrt{2\,\mathcal{M}}}\frac{1}{\left|\sqrt{\sinh^2(2\mathrm{r}) + 2|\alpha|^2e^{2\mathrm{r}}}\,d \psi_{\textrm{v},k}/dr_{\textrm{S}}\right| }\,,\nonumber\\
 &=  \frac{2\sqrt{2}}{\sqrt{\mathcal{M}}}\frac{1}{\sqrt{\sinh^2(2\mathrm{r}) + 2|\alpha|^2e^{2\mathrm{r}}}}\frac{{\bar{r}_0^2}}{ck\pi\tau_0}\frac{1}{|\chi|}
\end{align}
The error is independent of the cavity length, and it grows quadratically with $\bar{r}_0$. The error is smaller for larger times $\tau_0$. Therefore, integrating over long times yields higher precisions. As with proper time, classical states reach the normal scaling while the squeezed vacuum improves the precision by reaching the Heisenberg scaling.  The error is minimized for $|\chi|=1$.

\section{Discussion \label{sec:discussion}} 

\subsection{
Deformation of realistic rods
\label{sec:deformation}}
The previous analysis assumes that the rod that supports the clock mirrors is infinitely rigid. However, in practice, rods have finite rigidity. In this section we study the change in length in vertical clocks due to the proper acceleration that is necessary for the cavity to remain at a constant radial coordinate $\bar{r}_0$. A full analysis including inertial and tidal forces for a general spacetime based on Fermi normal coordinates has been given in \cite{Raetzelfreq:2018}. For the present case, we assume that the rod is supported at $x'=0$, where the vertical cavity intersects with the horizontal cavity. Each segment of the vertical rod in positive $x$-direction will then be compressed by the combined stress due to all segments above this segment. In the negative $x$-direction, the stress is negative and each segment will be stretched due to the added stress imposed on it by all segments below it. Therefore, the resulting length of the rod and the position of the end mirrors will depend on the position of the support. Eventually, one finds for the frequency of the vertical cavity \cite{Raetzelfreq:2018},
\begin{align}\label{physical:frequencies:relation:realistic:rod}
\nonumber \omega_{\textrm{v},k}= &\, \omega_{\textrm{h},k}\left(1 - \left(\frac{c^2}{c_\rm{rod}^2}+1\right)\frac{a^{x}L}{2c^2}\chi \right)\\
= &\, \omega_{\textrm{h},k}\left(1 - \Lambda\chi\right),
\end{align}

where $c_\rm{rod}$ is the speed of sound in the rod, we have conveniently introduced
\begin{equation}\label{eq:Lambda}
	\Lambda := \left(\frac{c^2}{c_\rm{rod}^2}+1\right)\frac{r_{\textrm{S}}L}{4\bar{r}_0^2} \,,
\end{equation} 
and we used that the proper acceleration is \begin{eqnarray}
	a^{x}= \frac{c^2r_{\textrm{S}}}{2\bar{r}_0^2}\,,
\end{eqnarray}
to first order in $r_{\textrm{S}}/\bar{r}_0$.
If we consider an aluminium rod, for example, then we have $c^2/c_\rm{rod}^2\approx 10^9$, while if we consider carbyne (linear acetylenic carbon), the material with the highest known specific modulus, then we can achieve $c^2/c_\rm{rod}^2\approx 10^8$  \cite{Mingjie2013}. This means that, in all realistic situations, the first term in $\Lambda$ will dominate.

\subsection{Bounds based on state-of-the-art parameters}

Let us see how well we can measure proper time using the electromagnetic field in a cavity. We choose a cavity length $L = 20 \rm{cm}$. Furthermore, we can assume a maximal squeezing of $\mathrm{r}=1.7$ which is about $15$dB -- the highest squeezing achieved to date \cite{Vahlbruch2016} leading to $\sinh^2(2\mathrm{r})\approx 224$. Using the reflectivity of current state-of-the-art mirrors \cite{crd}, the largest time scale for photons is of the order of $\tau_{l}=70\mathrm{ns}$, which corresponds to the time that it takes to decrease the initial photon number $n_i$ to $n_i/e$ \cite{svelto}. We can assume a power of $1$MW in the cavity (which leads to a necessary size of the beam at the mirrors of the order of $1\,\rm{cm}$ \cite{Meng:05}) and a mode with a wavelength of $L/(\pi k)=500\,$nm, which leads to a maximal number of photons $N_p$ of the order of $10^{14}$,  implying $N_p\approx |\alpha|^2e^{2\mathrm{r}}\gg \sinh^2(2\mathrm{r})$ for a squeezed coherent state \footnote{Although Heisenberg scaling is not reached for these values (the best precision is achieved when all the resources are put in a single-mode squeezed coherent state \cite{Monras:2006}), it is experimentally easier in the case of light to boost the coherence strength instead of increasing the squeezing parameter.}. In this case, squeezing is not enhancing the sensitivity in comparison to a coherent state with the same photon number and equation \eqref{final:absolute:precision:bounds:coherent state} can be used to calculate the general error bound. With the above numbers, we find a relative error bound $\delta_{\textrm{h},k} (\tau_0)$ of the order of $10^{-15}/\sqrt{\mathcal{M}}$, where we set $\tau_0=\tau_l$. The shortest time interval that can be resolved with the light clocks in principle is given by the absolute error bound $\Delta_{\textrm{h},k} (\tau_0)$. We find a fundamental time resolution of the order of $8\times 10^{-22}\,\rm{s}$ for the numbers above, where we considered $\mathcal{M}=1$. For longer time intervals $\tau_0$, we consider $\mathcal{M}=\tau_0/\tau_{l}$. For $\tau_0 = 1\,\rm{s}$, we find for example $\delta_{\textrm{h},k} (\tau_0) =  3\times 10^{-19}$. 

 Note that, if the time measurement is performed through a phase measurement, these error bounds are only valid if the frequency of the cavity is known with sufficiently large relative precision. As the cavity is affected by noise, for example thermal noise in the rod fixing the mirrors, the practical uncertainty is much larger. For example, highly stable optical resonators achieve a relative frequency stability of the order of $ \delta(\omega_k) \sim 10^{-17}$  for integration times between $1-100\,\rm{s}$ \cite{Robinson:19}. The total relative error given by the combination of the error in frequency and the fundamental error can be estimated through propagation of uncertainty as $\{[\delta(\omega_k)]^2 + [\delta_{\textrm{h},k} (\tau_0)]^2\}^{1/2}$, and hence, for $[\delta_{\textrm{h},k} (\tau_0)]^2\ll [\delta(\omega_k)]^2$, leads to a relative error of the same size as  $\delta(\omega_k)$. 

 On the surface of the Earth, that is $\bar{r}_0\approx 6.37\times 10^6\,$m, for an aluminium rod of $20\,\rm{cm}$, $\Lambda$ is of the order $10^{-8}$. Therefore, it is a main systematic that has to be taken into account in the description of extended deformable light clocks. The relativistic contribution to $\Lambda$, gravitational redshift inside the cavity, is of the order $10^{-17}$.
This is of the same order of magnitude as the realistic bound due to frequency fluctuations of the cavity that we discussed above and two orders of magnitude larger than the fundamental bound that we identified for the parameters above. Hence, state-of-the-art technology is close to the regime in which, on the surface of the Earth, a vertically oriented extended light clock of $20\rm{cm}$ resolves gravitational time dilation on or below its own length scale. Then, proper time is not simply a property of the whole clock in contrast to the case of point-like clocks. Instead, a reference point for time measurements within the clock has to be specified as we have done in an operational way above.

\section{\label{sec:conclusions} Conclusions}

Each mode of the quantum field in an optical cavity can be used as a spatially extended clock. We analyzed these clocks in two different orientations, vertical and horizontal in Schwarzschild spacetime. This analysis applies to light clocks in the spacetime of the Earth, therefore, the predictions can be tested in an Earth or space laboratory. The model is also useful to study clocks in the spacetime of an eternal black hole. Our results show that the vertical light clock is more precise than the horizontal light clock when the center of the vertical light clock is further away from the central object of the Schwarzschild spacetime than the horizontal light clock. 
This difference is proportional to the acceleration that is necessary to compensate the gravitational attraction, and thus, decays with the inverse square of the distance to the central object.
 As the fundamental precision limit of clocks scales with the inverse of their frequency, this result can be completely explained by gravitational redshift. The difference in ticking rate between the different clocks can be used to measure the acceleration of the clocks, and thus, the Schwarzschild radius. We gave an expression for the corresponding fundamental precision limit. 

We found that the optical cavity clock of $20\rm{cm}$ length can measure time with a relative precision of up to  $\sim 10^{-19}$ in principle and $\sim 10^{-17}$ as a realistic bound for an existing experimental system after one second of integration time. The maximal time resolution of a light cavity clock for the parameters that we considered is $\sim 10^{-21}\, \mathrm{s}$. The relativistic effect of gravitational time dilation in the gravitational field of the Earth is of the order of $10^{-17}$ for a clock of $20\rm{cm}$. Once this precision is exceeded, the measured time cannot be interpreted as a universal proper time that corresponds to the whole clock. Instead, a region smaller than the clock has to specified, where proper time corresponds to the time shown by the clock. In particular, an optical cavity clock at a fixed radial distance from a massive object runs at the proper time defined at its center to leading order, which includes gravitational acceleration but neglects gravity gradients.

 In the case of cavity light clocks, the end mirrors define fixed boundary conditions and the middle of the cavity can be specified with certainty. If we consider a quantum system like an atomic clock, also its position will be a quantized degree of freedom and no unique proper time can be associated with the time shown by the clock. As atomic clocks are now close to the regime in which they are able to resolve the gravitational time dilation on the length scale of their atoms' wave packets \cite{bothwell2022resolving,zheng2022differential}, the accuracy of time measurement approaches a fundamental boundary that has to be taken seriously. We enter a regime in which we have to describe clocks using a set of equations compatible with both, general relativity and the quantum properties of the clock. 

\begin{acknowledgments}
We acknowledge Richard Howl, Luis Cort\'es Barbado, Maximilian Lock, Jorma Louko, H{\'e}ctor Fern{\'a}ndez Melendez, and David Edward Bruschi for useful comments and considerations. T.B. acknowledges funding from CONACYT under project code 261699/359033. I.F acknowledges that this project was made possible through the support of a donation by John Moussouris and the grant ‘Leaps in cosmology: gravitational wave detection with quantum systems’ (No. 58745) from the John Templeton Foundation. D.R. thanks the Humboldt Foundation for supporting his work by awarding him their Feodor Lynen Research Fellowship and  acknowledges funding by the Deutsche Forschungsgemeinschaft (DFG, German Research Foundation) under Germany’s Excellence Strategy – EXC-2123 QuantumFrontiers – 390837967 and by the Federal Ministry of Education and Research of Germany in the project “Open6GHub” (grant number: 16KISK016). This work was in part supported by the Anglo-Austrian Society. 
\end{acknowledgments}

\bibliography{main}
\bibliographystyle{apsrev}


\end{document}